\documentclass[aps,prd,twocolumn,floatfix,nofootinbib]{revtex4}
\usepackage{graphicx}
\usepackage{amsmath}
\usepackage{latexsym}
\usepackage{here}
\usepackage{array}
\usepackage{dcolumn}
\usepackage{bm}

\newcommand{\half}{{\textstyle\frac{1}{2}}}

\def\lsim{\mathrel{\rlap{\raise 2.5pt \hbox{$<$}}\lower 2.5pt\hbox{$\sim$}}}
\def\gsim{\mathrel{\rlap{\raise 2.5pt \hbox{$>$}}\lower 2.5pt\hbox{$\sim$}}}

\renewcommand{\Re}{{\rm Re\thinspace}}
\renewcommand{\Im}{{\rm Im\thinspace}}

\allowdisplaybreaks

\def\vecalpha{{\pmb\alpha}}
\def\GeV{{\text{GeV}}}

\begin{document}


\title{CP Violation, Stability and Unitarity of the 
Two Higgs Doublet Model
}


\author{Abdul Wahab El Kaffas}
\email[]{awkaffas@ift.uib.no}
\affiliation{Department of Physics and Technology, University of Bergen,
Postboks 7803, N-5020 Bergen, Norway}
\author{Odd Magne Ogreid}
\email[]{omo@hib.no}
\affiliation{Bergen University College, Bergen, Norway}
\author{Per Osland}
\email[]{per.osland@ift.uib.no}
\affiliation{Department of Physics and Technology, University of Bergen,
Postboks 7803, N-5020 Bergen, Norway}

\date{\today}

\begin{abstract}
The Two-Higgs-Doublet Model is considered, in its CP-non-conserving version.
It is shown quantitatively how vacuum stability and tree-level unitarity in
the Higgs-Higgs-scattering sector constrain the parameter space of the model.
In particular, at high values of $\tan\beta$, the model violates unitarity,
unless some of the Higgs bosons are heavy. In the regime of large CP violation
in the neutral-Higgs--$t$-quark sector, which requires $\tan\beta\lsim1$, the
Yukawa coupling parameter space (determined by the neutral-Higgs-sector
rotation matrix) is reasonably unconstrained. On the other hand, the
corresponding neutral-Higgs--$b$-quark sector allows for large CP violation at
$\tan\beta\gg1$.  However, here the model is more constrained: Significant CP
violation is correlated with a considerable splitting among the two heavier
Higgs bosons.
\end{abstract}

\pacs{}

\maketitle

\section{Introduction}\label{sec:I}
The Two-Higgs-Doublet Model (2HDM) has been proposed as an extension to the
Standard Model (SM), in part because it provides an additional mechanism for
CP violation \cite{Lee:1973iz,Weinberg:1976hu,Branco:1985aq,Accomando:2006ga}.
Various experimental observations impose non-trivial constraints on it.  For
example, the $B-\bar B$ oscillations
\cite{Abbott:1979dt,Urban:1997gw,Ball:2006xx} and $Z\to b\bar b$ decay width
\cite{Denner:1991ie} exclude low values of $\tan\beta$, whereas the $B\to
X_s\gamma$ rate \cite{Buras:1993xp} excludes values of the charged-Higgs mass,
$M_{H^\pm}$, below approximately 300~GeV \cite{Misiak:2006zs}.  Also, the
precise measurements at LEP of the so-called $\rho$ parameter constrain the
mass splitting in the Higgs sector, and force the masses to be not too far
from the $Z$ mass scale \cite{Bertolini:1985ia}.  While these individual
constraints are all well-known, we are not aware of any dedicated attempt to
combine them, other than those of \cite{Grant:1994ak,Cheung:2003pw}.

There are also theoretical consistency conditions.  In particular, for vacuum
stability, the potential has to be positive for large values of the fields
\cite{Deshpande:1977rw,ElKaffas:2006nt}.  We shall furthermore require the
Higgs--Higgs scattering amplitudes to satisfy perturbative unitarity
\cite{Kanemura:1993hm,Akeroyd:2000wc,Ginzburg:2003fe}.  Taken together, these
constraints dramatically reduce the allowed parameter space of the model. 

The unitarity conditions are traditionally phrased in terms of upper bounds on
the Higgs masses \cite{Lee:1977eg,Kanemura:1993hm}. The present paper is
devoted to a study of the vacuum stability (or positivity) and unitarity
conditions. These limits will here be seen in conjunction with the
CP-violating Yukawa couplings.  We will study how the CP-violating couplings
are constrained by the stability and unitarity constraints, for various Higgs
mass scenarios.  The combination with experimental constraints will be
considered elsewhere.

In our parameterization of the model, we emphasize the masses and mixing
angles. The latter are closely related to the Yukawa couplings, and thus
somewhat more ``physical'' than the parameters of the potential, to which they
are clearly related.

\section{The model}
\setcounter{equation}{0}
\label{sect:model}

The present study is limited to the 2HDM~(II), which is defined by having one
Higgs doublet ($\Phi_2$) couple to the up-type quarks, and the other
($\Phi_1$) to the down-type quarks \cite{HHG}.

We take the 2HDM potential to be para\-metrized as:
\begin{align}
\label{Eq:pot_7}
V&=\frac{\lambda_1}{2}(\Phi_1^\dagger\Phi_1)^2
+\frac{\lambda_2}{2}(\Phi_2^\dagger\Phi_2)^2
+\lambda_3(\Phi_1^\dagger\Phi_1) (\Phi_2^\dagger\Phi_2) \nonumber \\
&+\lambda_4(\Phi_1^\dagger\Phi_2) (\Phi_2^\dagger\Phi_1)
+\frac{1}{2}\left[\lambda_5(\Phi_1^\dagger\Phi_2)^2+{\rm h.c.}\right] \\
&-\frac{1}{2}\left\{m_{11}^2(\Phi_1^\dagger\Phi_1)
\!+\!\left[m_{12}^2 (\Phi_1^\dagger\Phi_2)\!+\!{\rm h.c.}\right]
\!+\!m_{22}^2(\Phi_2^\dagger\Phi_2)\right\} \nonumber
\end{align}
Thus, the $Z_2$ symmetry will be respected by the quartic terms, and
Flavour-Changing Neutral Currents are constrained \cite{Glashow:1976nt}.  We
shall refer to this model (without the $\lambda_6$ and $\lambda_7$ terms) as
the $\text{2HDM}_5$. The more general model, with also $\lambda_6$ and
$\lambda_7$ couplings, will be discussed elsewhere.

We allow for CP violation, i.e., $\lambda_5$ and $m_{12}^2$ may be complex.
All three neutral states will then mix,
\begin{equation}
\label{Eq:cal-M}
R{\cal M}^2R^{\rm T}={\cal M}^2_{\rm diag}={\rm diag}(M_1^2,M_2^2,M_3^2),
\end{equation}
where ${\cal M}^2$ is determined from second derivatives of the above
potential.  The $3\times3$ mixing matrix $R$ governing the neutral sector will
be parametrized in terms of the angles $\alpha_1$, $\alpha_2$ and $\alpha_3$
as in \cite{Accomando:2006ga,Khater:2003wq}:
\begin{equation}
R=
\begin{pmatrix}
c_1\,c_2 & s_1\,c_2 & s_2 \\
- (c_1\,s_2\,s_3\!+\!s_1\,c_3) 
& c_1\,c_3\!-\!s_1\,s_2\,s_3 & c_2\,s_3 \\
- c_1\,s_2\,c_3\!+\!s_1\,s_3 
& - (c_1\,s_3\!+\!s_1\,s_2\,c_3) & c_2\,c_3
\end{pmatrix}
\end{equation}
where $c_1=\cos\alpha_1$, $s_1=\sin\alpha_1$, etc., and
\begin{equation}
-\frac{\pi}{2}<\alpha_1\le\frac{\pi}{2},\quad
-\frac{\pi}{2}<\alpha_2\le\frac{\pi}{2},\quad
0\le\alpha_3\le\frac{\pi}{2}.
\end{equation}
(In ref.~\cite{Khater:2003wq}, the angles are denoted as
$\tilde\alpha=\alpha_1$, $\alpha_b=\alpha_2$, $\alpha_c=\alpha_3$.)  For these
angular ranges, we have $c_i\ge0$, $s_3\ge0$, whereas $s_1$ and $s_2$ may be
either positive or negative.  We will use the terminology ``general 2HDM'' as
a reminder that CP violation is allowed.

\begin{figure}[hbt]
\vspace*{-5mm}
\begin{center}
\includegraphics[width=7cm]{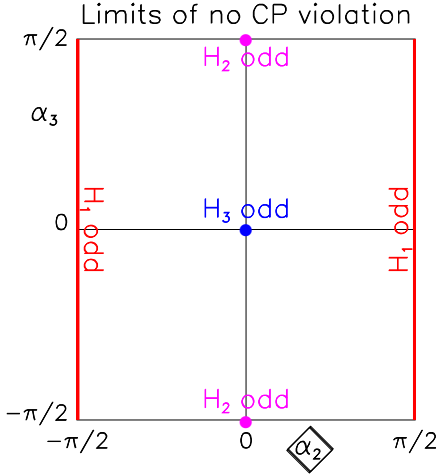}
\vspace*{-5mm}
\caption{\label{Fig:alpha23-scheme} Limits of no CP-violation shown in the
$\alpha_2$--$\alpha_3$ plane, with the CP-odd Higgs boson identified.}
\end{center}
\vspace*{-2mm}
\end{figure}

With all three masses different, there are three limits of {\it no} CP
violation, i.e., with two Higgs bosons that are CP even and one that is odd.
The three limits are \cite{ElKaffas:2006nt}: 
\begin{alignat}{2} \label{Eq:CP-cons}
&\text{$H_1$ odd:} &\quad
&\alpha_2\simeq\pm\pi/2,\ \alpha_1, \alpha_3 \text{ arbitrary}, \nonumber \\
&\text{$H_2$ odd:} &\quad
&\alpha_2=0,\ \alpha_3=\pi/2,\ \alpha_1 \text{ arbitrary}, \nonumber \\
&\text{$H_3$ odd:} &\quad
&\alpha_2=\alpha_3=0,\ \alpha_1 \text{ arbitrary}.
\end{alignat}
These limits of no CP-violation are indicated in
Fig.~\ref{Fig:alpha23-scheme}.  (For future reference, we display in
Fig.~\ref{Fig:alpha23-scheme} the full range $-\pi/2<\alpha_3\le\pi/2$, as is
required for the general case of non-zero $\lambda_6$ and $\lambda_7$.)

In the general CP-non-conserving case, the neutral sector is conveniently
described by these three mixing angles, together with two masses $(M_1, M_2)$,
$\tan\beta$ and the parameter $\mu^2=\Re m_{12}^2/(2\cos\beta\sin\beta)$.  In
fact, since the Yukawa couplings are compactly expressed in terms of these
rotation matrix elements,
\begin{alignat}{2}  \label{Eq:H_itt}
&H_j  b\bar b: &\qquad 
&\frac{1}{\cos\beta}\, [R_{j1}-i\gamma_5\sin\beta R_{j3}], \nonumber \\
&H_j  t\bar t: &\qquad 
&\frac{1}{\sin\beta}\, [R_{j2}-i\gamma_5\cos\beta R_{j3}],
\end{alignat}
the angles provide a rather physical way to parametrize the model.

From Eq.~(\ref{Eq:cal-M}), it follows that
\begin{equation}
\label{Eq:calM-RMsqR}
({\cal M}^2)_{ij}=\sum_k R_{ki} M_k^2 R_{kj}.
\end{equation}
In the general CP-non-conserving case, both $({\cal M}^2)_{13}$ and $({\cal
M}^2)_{23}$ will be non-zero. In fact, they are related by
\begin{equation} \label{Eq:rel13-23}
({\cal M}^2)_{13}=\tan\beta({\cal M}^2)_{23}.
\end{equation}
From these two equations, (\ref{Eq:calM-RMsqR}) and (\ref{Eq:rel13-23}), we
can determine $M_3$ from $M_1$, $M_2$, the angles
$\vecalpha=(\alpha_1,\alpha_2,\alpha_3)$ and $\tan\beta$ \cite{Khater:2003wq}:
\begin{equation} \label{Eq:M_3}
M_3^2=\frac{M_1^2R_{13}(R_{12}\tan\beta\!-\!R_{11})
\!+\!M_2^2R_{23}(R_{22}\tan\beta\!-\!R_{21})}
{R_{33}(R_{31}-R_{32}\tan\beta)}
\end{equation}
where we impose $M_1\le M_2\le M_3$.

\section{Extracting the $\lambda$}
\setcounter{equation}{0}
\label{sect:lambdas}

As discussed above,
in the $\text{2HDM}_5$, with $\Im\lambda_5\ne0$, the two masses $M_1$ and
$M_2$ will together with $\vecalpha$ and $\tan\beta$ determine $M_3$.
Providing also $M_{H^\pm}$ and $\mu^2$, all the $\lambda$'s are determined as
follows.  Since the left-hand side of (\ref{Eq:calM-RMsqR}) can be expressed
in terms of the parameters of the potential (see, for example,
\cite{ElKaffas:2006nt}), we can solve these equations and obtain the
$\lambda$'s in terms of the rotation matrix, the neutral mass eigenvalues,
$\mu^2$ and $M_{H^\pm}$:
\begin{align} \label{Eq:lambda1}
\lambda_1&=\frac{1}{c_\beta^2v^2}
[c_1^2c_2^2M_1^2
+(c_1s_2s_3+s_1c_3)^2M_2^2 \nonumber \\
&+(c_1s_2c_3-s_1s_3)^2M_3^2
-s_\beta^2\mu^2], \\
\lambda_2&=\frac{1}{s_\beta^2v^2} \label{Eq:lambda2}
[s_1^2c_2^2M_1^2
+(c_1c_3-s_1s_2s_3)^2M_2^2 \nonumber \\
&+(c_1s_3+s_1s_2c_3)^2M_3^2-c_\beta^2\mu^2], \\
\lambda_3&=\frac{1}{c_\beta s_\beta v^2} \label{Eq:lambda3}
\{c_1s_1[c_2^2M_1^2+(s_2^2s_3^2-c_3^2)M_2^2 \nonumber \\
&+(s_2^2c_3^2-s_3^2)M_3^2]
+s_2c_3s_3(c_1^2-s_1^2)(M_3^2-M_2^2)\} \nonumber \\
+&\frac{1}{v^2}[2M_{H^\pm}^2-\mu^2], \\
\lambda_4&=\frac{1}{v^2}
[s_2^2M_1^2+c_2^2s_3^2M_2^2+c_2^2c_3^2M_3^2
+\mu^2-2M_{H^\pm}^2], \\
\Re\lambda_5&=\frac{1}{v^2}
[-s_2^2M_1^2-c_2^2s_3^2M_2^2-c_2^2c_3^2M_3^2+\mu^2], \\
\Im\lambda_5&=\frac{-1}{c_\beta s_\beta v^2}
\{c_\beta[c_1c_2s_2M_1^2-c_2s_3(c_1s_2s_3+s_1c_3)M_2^2 \nonumber \\
&+c_2c_3(s_1s_3-c_1s_2c_3)M_3^2] 
+s_\beta[s_1c_2s_2M_1^2 \label{Eq:lambda5I} \\
&+c_2s_3(c_1c_3\!-\!s_1s_2s_3)M_2^2
\!-\!c_2c_3(c_1s_3\!+\!s_1s_2c_3)M_3^2]\}, \nonumber 
\end{align}
where $c_\beta=\cos\beta$, $s_\beta=\sin\beta$.

While $M_3^2$ is given in terms of $M_1^2$, $M_2^2$, $R$ and $\tan\beta$ by
Eq.~(\ref{Eq:M_3}), it is more transparent not to substitute for $M_3^2$ in
these expressions (\ref{Eq:lambda1})--(\ref{Eq:lambda5I}).  These equations
are the analogues of those of \cite{Casalbuoni:1987cz} for the CP-conserving
$\text{2HDM}_5$.
\subsection{Large values of $\mu^2$}
At large $\mu^2\gg M_3^2$, it is seen from (\ref{Eq:lambda1}) and
(\ref{Eq:lambda2}) that $\lambda_1$ and $\lambda_2$ will eventually turn
negative. This would violate stability and the model would break down.  Thus,
for fixed Higgs masses, there is an upper limit to $\mu^2$.

\subsection{Large values of $\tan\beta$}
\label{sect:largetanbeta}
According to the constraints of unitarity, reviewed in
Sec.~\ref{sect:unitarity}, the couplings $\lambda_1$, $\lambda_2$ and
$|\lambda_3|$ cannot be too large.  At large values of $\tan\beta$, where
$c_\beta\equiv\cos\beta\to0$, the coefficients in Eq.~(\ref{Eq:lambda1})
multiplying $M_2^2$ and $M_3^2$ will hence be constrained.  When $\mu^2$ is
small, these coefficients must be small.  This requires $|s_1|$ and $|s_2|$
both to be small. Otherwise, when $\mu^2$ is relevant, the terms proportional
to $M_2^2$ and $M_3^2$ must balance against the $\mu^2$-term.

\section{Positivity and unitarity}
\setcounter{equation}{0}
\label{sect:general}

We shall project the constraints of positivity and unitarity onto the
$\tan\beta$--$M_{H^\pm}$ plane.  Such a projection of information from a
six-dimensional space onto a point in the $\tan\beta$--$M_{H^\pm}$ plane can
be done in a variety of ways, all of which will lead to some loss of
information.  However, we feel that this loss of detailed information can be
compensated for by the ``overview'' obtained by the following procedure:
\begin{enumerate}
\item
Pick a set of neutral-Higgs-boson masses, 
$(M_1, M_2)$ together with $\mu^2$.
\item
Scan an $N=n_1\times n_2\times n_3$ grid in
the $(\alpha_1,\alpha_2,\alpha_3)$ space, and count
the number $j$ of these points that give a viable model.
(Alternatively, one could scan over $N$ random points in this space.)
\item
The ratio
\begin{equation} \label{Eq:neutr_Q}
Q=j/N, \quad 0\le Q\le 1,
\end{equation}
is then a figure of merit, a measure of ``how allowed'' the point is, in the
$\tan\beta$--$M_{H^\pm}$ plane.  If $Q=0$, no sampled point in the
$\vecalpha=(\alpha_1,\alpha_2,\alpha_3)$ space is allowed. Similarly, if
$Q=1$, they are all allowed.  An alternative measure
\begin{equation} \label{Eq:neutr_Q0}
Q_+=j/N_+,\quad Q_+\ge Q,
\end{equation}
counts in the denominator
only those points $N_+$ for which positivity is satisfied.
\end{enumerate}

Of course the 2HDM, if realized in nature, would only exist at {\it one} point
in this parameter space.  However, we think the above quantities $Q$ and $Q_+$
give meaningful measures of how ``likely'' different parameters are.

\subsection{Reference masses}
\label{sect:refmasses}
We shall impose the conditions of positivity and unitarity on the model, for
the different ``reference'' mass sets given in Table~\ref{tab:refmasses} (and
variations around these).  For each of these mass sets we scan the model
properties in the $\vecalpha$ space.  From these reference masses, some trends
will emerge, allowing us to draw more general conclusions.
\begin{table}[ht]
\begin{center}
\renewcommand{\tabcolsep}{.75em}
\begin{tabular}{|c|c|c|c|c|}
\hline 
Name&$M_1 [\GeV]$&$M_2 [\GeV]$&$\mu^2~[\GeV]^2$\\
\hline
``100-300''&100&300&0 [$\pm(200)^2$]\\
``150-300''&150&300&0 [$\pm(200)^2$]\\
``100-500''&100&500&0 [$\pm(200)^2$]\\
``150-500''&150&500&0 [$\pm(200)^2$]\\
\hline
\end{tabular} 
\caption{Reference masses.}\label{tab:refmasses}
\end{center}
\vspace*{-2mm}
\end{table}

These masses are inspired by the indication from LEP that there is a
relatively light Higgs boson \cite{unknown:2005di}, here denoted $H_1$.  The
others, $H_2$ and $H_3$, are then presumably more massive, and do not directly
affect the LEP phenomenology.  As an alternative, we shall also briefly
consider the case of {\it two} light Higgs bosons, with the third one
considerably more massive (see Sec.~\ref{sect:two-light}).

\subsection{Stability}

Let us first explore the effect of imposing vacuum stability, or positivity.
The positivity conditions can be formulated as (for a general discussion, see
Appendix~A of \cite{ElKaffas:2006nt}):
\begin{align} \label{Eq:positivity}
\lambda_1&>0, \quad \lambda_2>0, \nonumber \\
\lambda_3&+\min[0,\lambda_4-|\lambda_5|]>-\sqrt{\lambda_1\lambda_2}.
\end{align}
Actually, we will use the notion ``positivity'' to include also the
non-trivial conditions $M_3^2>0$ and $M_2\le M_3$ [see Eq.~(\ref{Eq:M_3})].
We scan the $\vecalpha$ parameter space as discussed above, and show in
Table~\ref{tab:positivity} the fraction of parameter points that satisfy
``positivity'', as defined above.

\begin{table}[ht]
\begin{center}
\renewcommand{\tabcolsep}{.75em}
\begin{tabular}{|c|c|c|c|}
\hline 
$\mu^2$&$-(200~\text{GeV})^2$&$0$&$(200~\text{GeV})^2$\\
\hline
``100-300''&30.8--31.0\%&30.8--31.0\%&24.0--28.0\%\\
``150-300''&30.8--31.0\%&30.8--31.0\%&26.2--31.0\%\\
``100-500''&30.8--31.0\%&30.8--31.0\%&27.3--29.7\%\\
``150-500''&30.8--31.0\%&30.8--31.0\%&28.8--31.0\%\\
\hline
\end{tabular} 
\caption{Percentage $Q$ of points in $\vecalpha$ space for which
positivity is satisfied.}\label{tab:positivity}
\end{center}
\vspace*{-2mm}
\end{table}

We shall henceforth refer to the set of points in the $\vecalpha$ space where
positivity is satisfied, as $\vecalpha_+$.  The fraction $Q$ of points in the
$\vecalpha$ space for which positivity is satisfied, is around 30\%.  (The
range given indicates the lowest and highest values found when scanning over
$0.5\le\tan\beta\le50$ and $200~\text{GeV}\le M_{H^\pm}\le 700~\text{GeV}$.)
We note that an upper bound for this fraction is 50\%.  This comes about from
the fact that for a given value of $\beta+\alpha_1$, in the $\text{2HDM}_5$,
only positive or only negative values of $\alpha_2$ are allowed, not both
\cite{Khater:2003wq}.  Thus, with $0\le\alpha_3\le\pi/2$, the sign of
$\alpha_2$ will be given by that of $\beta+\alpha_1$:
\begin{alignat}{2}
0&<\beta+\alpha_1<\half\pi: &\quad &\alpha_2\alpha_3>0, \nonumber \\
-\half\pi&<\beta+\alpha_1<0: &\quad &\alpha_2\alpha_3<0.
\end{alignat}
For small and negative values of $\mu^2$, ``most'' of the exclusion provided
by the positivity constraint is already contained in the conditions $M_3^2>0$
and $M_3>M_2$, without the explicit conditions (\ref{Eq:positivity}) on the
$\lambda$'s.  The conditions on the $\lambda$'s provide the additional
exclusion at positive values of $\mu^2$ that is evident in
Table~\ref{tab:positivity}.

\subsection{Unitarity}
\label{sect:unitarity}

Perturbative unitarity in the Higgs--Higgs sector imposes upper bounds on the
$|\lambda_i|$. These relations have the structure
\begin{equation}
\frac{1}{16\pi}\big|\sum a_i\lambda_i+\sqrt{Q(\lambda_i)}\big|\le1
\end{equation}
where the coefficients $a_i$ are of ${\cal O}(1)$ and $Q(\lambda_i)$ is a
quadratic expression \cite{Kanemura:1993hm,Akeroyd:2000wc,Ginzburg:2003fe}.
These constraints can be expressed as upper bounds on the Higgs masses
\cite{Kanemura:1993hm}.  They are conveniently formulated in terms of the
different weak isospin and hypercharge channels \cite{Ginzburg:2003fe}.

\begin{figure}[t!]
\vspace*{-5mm}
\begin{center}
\includegraphics[width=9cm]{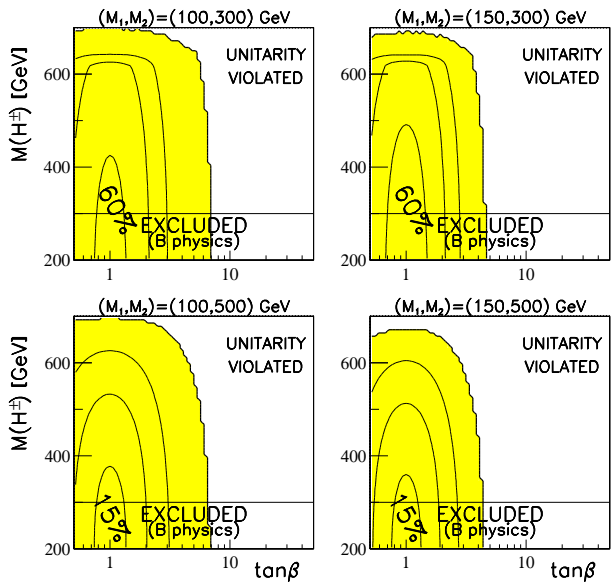}
\vspace*{-10mm}
\caption{\label{Fig:unitarity-musq=0} Percentage of points $Q_+$ in the
$\vecalpha_+$ space that satisfy unitarity.  Four sets of $(M_1,M_2)$ values
are considered, as indicated.  All panels: $\mu^2=0$. Yellow region: $Q_+>0$
(positivity and unitarity satisfied). The contours show $Q_+=0$, 20\%, 40\%
and 60\% (upper panels) and 0, 5\%, 10\%  and 15\% (lower panels).  The
line at $M_{H^\pm}=300$~GeV indicates roughly what is excluded by the $b\to
s\gamma$ constraint \cite{Misiak:2006zs}.}
\end{center}
\vspace*{-2mm}
\end{figure}

When $\mu^2=0$ (or negative), we see from Eq.~(\ref{Eq:lambda1}) that
$\lambda_1$ will become large when $\tan\beta\gg1$. Thus, unitarity will at
some point be violated. This is illustrated in
Fig.~\ref{Fig:unitarity-musq=0}, where we show in yellow where at least one
point in $\vecalpha_+$ satisfies positivity and unitarity.

\section{Higgs-mediated CP violation}
\label{sec:CP-violation}
\setcounter{equation}{0}

When the Yukawa couplings contain both a scalar and a pseudoscalar term,
as in Eq.~(\ref{Eq:H_itt}), the exchange of Higgs particles leads to
CP violation via an amplitude, which for couplings to 
$b$ and $t$-quarks is proportional to
\begin{equation} \label{Eq:Y-CP-b}
Y^b_\text{CP}=\sum_{j=1}^3 R_{j1}R_{j3}\,\frac{\sin\beta}{\cos^2\beta}\,
f_b(M_j^2)
\end{equation}
and
\begin{equation} \label{Eq:Y-CP-t}
Y^t_\text{CP}=\sum_{j=1}^3 R_{j2}R_{j3}\,\frac{\cos\beta}{\sin^2\beta}\,
f_t(M_j^2),
\end{equation}
respectively.
The function $f_q(M_j^2)$ is in general some loop integral that depends
on the Higgs mass $M_j$.
This CP-violating effect is most important when the Higgs masses
are not too close. Otherwise there are cancellations among 
different contributions, due to the orthogonality of the rotation matrix, 
\begin{equation} \label{Eq:orthogonal}
\sum_{j=1}^3 R_{j1}R_{j3}=0, \quad
\sum_{j=1}^3 R_{j2}R_{j3}=0.
\end{equation}
Also, since the functions $f_q(M^2)$ decrease for high
values of $M^2$, the effect tends to be larger
when the lightest Higgs boson is reasonably light.

Let us therefore focus on the couplings of the lightest Higgs boson, $H_1$.
For maximal CP violation in the $H_1 b\bar b$ coupling,
$R_{11}R_{13}\sin\beta/\cos^2\beta=\half\cos\alpha_1 
\sin2\alpha_2\sin\beta/ \cos^2\beta$ must be large, requiring
\begin{equation} \label{Eq:H_1bb-maxcpv}
\alpha_1\simeq0, \quad
\alpha_2\simeq\pm\pi/4, \quad
\tan\beta\gg1.  
\end{equation}
Similarly, for
maximal CP violation in the $H_1 t\bar t$ coupling,
$R_{12}R_{13}\cos\beta/\sin^2\beta=\half
\sin\alpha_1\sin2\alpha_2\cos\beta/\sin^2\beta$ must be large, or
\begin{equation} \label{Eq:H_1tt-maxcpv}
\alpha_1\simeq\pm\pi/2, \quad
\alpha_2\simeq\pm\pi/4, \quad
\tan\beta\lsim1.
\end{equation}

When the two heavier Higgs bosons have a similar mass,
$M=M_2\simeq M_3$,
the expression (\ref{Eq:Y-CP-t}), for example,
simplifies because of (\ref{Eq:orthogonal}):
\begin{equation} \label{Eq:Y-CP-t-approx}
Y^t_\text{CP}\simeq R_{12}R_{13}\,\frac{\cos\beta}{\sin^2\beta}
\bigl[f_t(M_1^2)-f_t(M^2)\bigr].
\end{equation}
Thus, also in this case is the coupling of the lightest Higgs boson of special
importance.

These two conditions (\ref{Eq:H_1bb-maxcpv}) and (\ref{Eq:H_1tt-maxcpv}) will
be studied in the following.  Common to both of them is the requirement that
$\alpha_2\simeq\pm\pi/4$.  Vice versa, there is no CP violation mediated by
the lightest Higgs exchange when $\alpha_2\simeq0$ or when
$\alpha_2\simeq\pm\pi/2$.  Also, we note that the conditions
(\ref{Eq:H_1bb-maxcpv}) and (\ref{Eq:H_1tt-maxcpv}) do not refer to
$\alpha_3$.

In the case when {\it two} Higgs bosons are fairly light compared to the third
one, by orthogonality, it will be the couplings of the heavy one that
determine the amount of CP violation.  For maximal CP violation in the
$H_3b\bar b$ coupling, $|R_{31}R_{33}|\sin\beta/\cos^2\beta
=|-c_1s_2c_3+s_1s_3|c_2c_3\sin\beta/\cos^2\beta$ must then be large, requiring
$c_2$ and $c_3$ to be non-zero. More explicitly, for small $|\alpha_1|$ one
must have $\alpha_2\simeq\pm\pi/4$ and $\alpha_3\simeq0$.  At the other
extreme, for $|\alpha_1|\simeq\pi/2$, one must have $\alpha_2\simeq0$ and
$\alpha_3\simeq\pi/4$.  Similarly, for maximal CP violation in the $H_3t\bar
t$ coupling, $|R_{32}R_{33}|\cos\beta/\sin^2\beta
=|c_1s_3+s_1s_2c_3|c_2c_3\cos\beta/\sin^2\beta$ must be large, also requiring
$c_2$ and $c_3$ to be non-zero. More explicitly, for small $|\alpha_1|$ one
must have $\alpha_2\simeq0$ and $\alpha_3\simeq\pi/4$.  At the other
extreme, for $|\alpha_1|\simeq\pi/2$, one must have $\alpha_2\simeq\pm\pi/4$
and $\alpha_3\simeq0$.

We shall now proceed to study to what extent these regions of large
CP-violation are allowed by the stability and unitarity constraints.

\section{Allowed regions for $\mu=0$}
\label{sect:mu=0}
\setcounter{equation}{0}

The unitarity constraint can have a rather dramatic effect at ``large'' values
of $\tan\beta$ and $M_{H^\pm}$.  While the general constraints on the
charged-Higgs sector exclude low values of $\tan\beta$ and $M_{H^\pm}$ (see
\cite{Abbott:1979dt,Urban:1997gw,Ball:2006xx,Denner:1991ie,Buras:1993xp,
Misiak:2006zs,Abulencia:2006mq,Yao:2006px}), the constraints of unitarity
exclude high values of these same parameters.  Only some region in the middle
remains not excluded.  For $(M_1,M_2)=(100,300)$~GeV and $\mu=0$ [see
Fig.~\ref{Fig:unitarity-musq=0}], unitarity excludes everything above
$\tan\beta\sim7$ (for any value of $M_{H^\pm}$), and above
$M_{H^\pm}\sim700$~GeV (for any value of $\tan\beta$).

\begin{figure}[t!]
\vspace*{-5mm}
\begin{center}
\includegraphics[width=9.0cm]{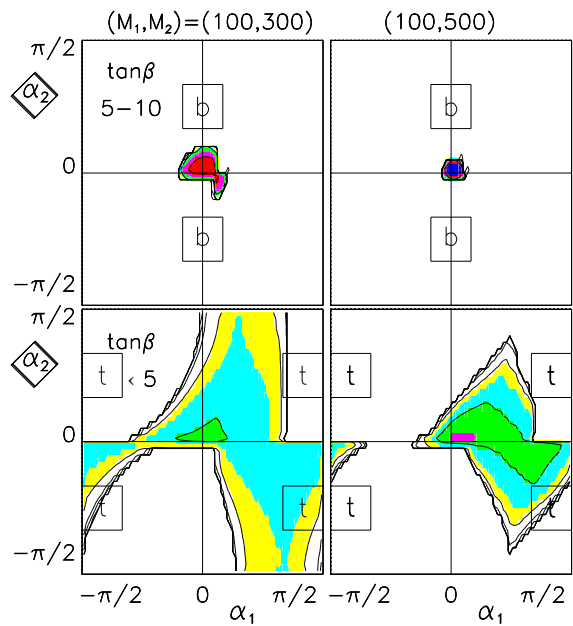}
\vspace*{-5mm}
\caption{\label{Fig:alphas-cont2-300-500} Allowed regions in the
$\alpha_1$--$\alpha_2$ plane, for $(M_1,M_2)=(100,300)~\GeV$ and
$(100,500)~\GeV$, $\mu=0$ and two slices in $\tan\beta$ as indicated.  At
higher values of $\tan\beta$ there are no allowed points (see also
Fig.~\ref{Fig:unitarity-musq=0}).  Contours are shown at each negative power
of 10, as appropriate.  Yellow (light blue) indicates where the normalized
distribution is higher than $10^{-4}$ ($3\times10^{-4}$); green (purple)
levels above $10^{-3}$ ($3\times10^{-3}$); red (dark blue) is above $10^{-2}$
($3\times10^{-2}$).
Regions of major CP violation are labeled ``$b$'' and ``$t$''.}
\end{center}
\vspace*{-2mm}
\end{figure}

For $M_2=300$~GeV, the percentage of points in $\vecalpha_+$ space for which
unitarity is satisfied, reaches (at low $\tan\beta$ and low $M_{H^\pm}$)
beyond 60\%, whereas for $M_2=500$~GeV, it only reaches values close to
20\%.

The domains in which solutions exist ($Q_+>0$) depend on $\mu^2$: For negative
values of $\mu^2$, the region typically shrinks to lower values of
$\tan\beta$, for positive values of $\mu^2$ it extends to larger values of
$\tan\beta$ (see next section).  However, the maximum values of $Q_+$ (at low
values of $\tan\beta$), are little changed.

\begin{figure}[b!]
\begin{center}
\vspace*{-2mm}
\includegraphics[width=9.0cm]{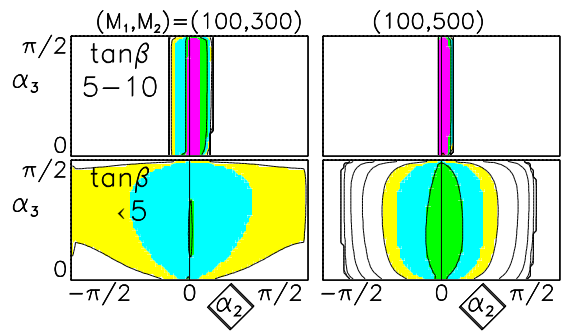}
\vspace*{-5mm}
\caption{\label{Fig:alphas-cont2-300-500-23}
Allowed regions in the $\alpha_2$--$\alpha_3$ plane, for
$(M_1,M_2)=(100,300)~\GeV$ and $(100,500)~\GeV$, $\mu=0$
and two slices in $\tan\beta$ as indicated.
Contours and colour coding as in Fig.~\ref{Fig:alphas-cont2-300-500}.}
\end{center}
\vspace*{-2mm}
\end{figure}

As discussed in Sect.~\ref{sect:largetanbeta}, for large values of
$\tan\beta$, the allowed solutions get constrained to a region of $|\alpha_1|$
and $|\alpha_2|$ both small. 
This is illustrated in
Fig.~\ref{Fig:alphas-cont2-300-500}, where we show regions of allowed
$\alpha_1$ and $\alpha_2$, for $\tan\beta<5$ (lower part) and $5<\tan\beta<10$
(upper part). The masses considered are $(M_1,M_2)=(100,300)~\GeV$ and
$(100,500)~\GeV$. No allowed solutions were found for $\tan\beta\gsim7$. As
$M_2$ is increased from 300~GeV to 500~GeV, the allowed region shrinks.

As discussed in Sec.~\ref{sec:CP-violation}, the Yukawa couplings of the
lightest Higgs particle to $b$ and $t$ quarks is large for $\tan\beta$ low or
high, with $|\alpha_1|\simeq\pi/2$ or 0, respectively, and
$|\alpha_2|\simeq\pi/4$ in both cases. These regions will be referred to as
``regions of major CP violation'' and are indicated by boxes labeled ``$b$''
and ``$t$'' in Fig.~\ref{Fig:alphas-cont2-300-500}.  We note that the boxes
labeled ``$b$'' are empty, the model does not give large CP violation in the
$bbH_1$ couplings for these mass and $\mu$ parameters.

Since at high $\tan\beta$, $|\alpha_1|$ and $|\alpha_2|$ are small, $M_3$ will
be almost degenerate with $M_2$.  The distribution of $M_3$ values is shown in
Table~\ref{tab:mh3-dist-mu=0}.  It is seen that $M_3$ is just barely larger
than $M_2$, in particular for high values of $\tan\beta$ and high values of
$M_2$ (cf.\ $M_2=500$~GeV vs.\ 300~GeV).

\begin{table}[htb]
\begin{center}
\renewcommand{\tabcolsep}{.75em}
$(M_1,M_2)=(100,300\ [500])$~GeV
\begin{tabular}{|c|c|c|c|}
\hline 
$\tan\beta$&$\xi<1.1$&$1.1<\xi<1.5$&$1.5<\xi$\\
\hline
$5-10$&94.4\ [98.5]\%&5.4\ [\ 1.5]\%&0.2\ [\ 0.0]\%\\
$<5$&41.1\ [82.5]\%&50.2\ [17.5]\%&8.6\ [\ 0.0]\%\\
\hline
\end{tabular} 
\caption{Distribution of $M_3$ values, $\xi=M_3/M_2$.
Contours and colour coding as in Fig.~\ref{Fig:alphas-cont2-300-500}.}
\label{tab:mh3-dist-mu=0}
\end{center}
\vspace*{-5mm}
\end{table}

The distribution in $\alpha_3$, of allowed solutions, is more spread out, 
as shown in Fig.~\ref{Fig:alphas-cont2-300-500-23}.

At large values of $\tan\beta$ it turns out to be the isospin-zero,
hypercharge-zero channel that is most constraining.

When $\mu^2<0$, the range in $\tan\beta$ is likewise limited, and the allowed
regions in $\alpha_1$, $\alpha_2$ and $\alpha_3$ are similar to those for the
$\mu^2=0$ case.

\section{Allowed regions for $ M_1\lsim \mu$}
\setcounter{equation}{0}
The large-$\mu$ case is often referred to as the decoupling limit.  It has
received considerable attention in the CP-conserving case
\cite{Gunion:2002zf}.  Within the framework set up by our choice of input
parameters, it is natural to distinguish three mass scales: $M_1$, $M_2$ and
$\mu$.  Thus, there are three cases: 
\begin{alignat}{3}
(i)& &\quad &\mu<M_1<M_2, &\quad & \text{Sect.\ref{sect:mu=0}}, \nonumber \\
(ii)& &\quad &M_1<\mu<M_2, &\quad & \text{decoupling}, \nonumber \\
(iii)& &\quad &M_1<M_2<\mu, &\quad & \text{decoupling}. 
\end{alignat}
If $\mu$ is ``significantly'' larger than $M_1$, the latter two both
correspond to decoupling in the sense of Gunion and Haber
\cite{Gunion:2002zf}, but from the point of view of CP violation, they can be
rather different.

\begin{figure}[t!]
\begin{center}
\vspace*{-5mm}
\includegraphics[width=9.0cm]{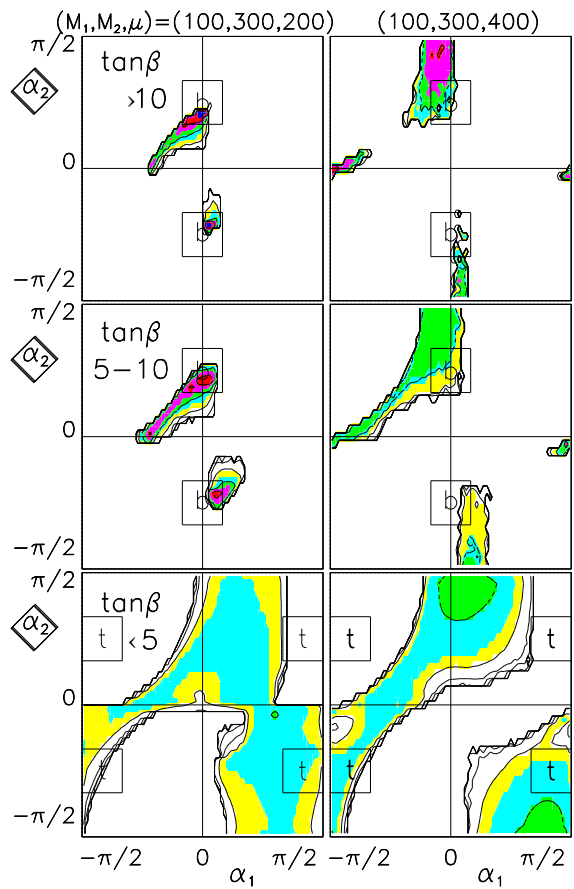}
\vspace*{-8mm}
\caption{\label{Fig:alphas-300} Allowed regions in the $\alpha_1$--$\alpha_2$
plane, for $(M_1,M_2)=(100,300)~\GeV$, $\mu=200$~GeV and 400~GeV and three
slices in $\tan\beta$ as indicated.  Contours and colour coding as in
Fig.~\ref{Fig:alphas-cont2-300-500}.  Regions of major CP violation are
labeled ``$b$'' and ``$t$''.}
\end{center}
\vspace*{-4mm}
\end{figure}

In these regimes of $M_1<\mu$, it is possible to keep $\lambda_1$ and
$|\lambda_3|$ within the allowed range (not too large) by carefully tuning the
other parameters.  For $\mu$ suitably chosen (large), no part of the
$\tan\beta$--$M_{H^\pm}$ plane is disallowed. From an inspection of
Eq.~(\ref{Eq:lambda1}) for $\lambda_1$, we see that $|s_1|$ and/or $|s_2|$
must be small.  But they can not both be zero, unless $M_1$ is very close to
$\mu$.  This region of small $|s_1|$ and/or $|s_2|$ will also yield solutions
for $\lambda_3$ that are sufficiently small.  In the following, we discuss the
specific examples of $(M_1,M_2)=(100,300)$~GeV and (100,500)~GeV, each of them
for two values of $\mu$.

\begin{figure}[b!]
\begin{center}
\vspace*{-2mm}
\includegraphics[width=9.0cm]{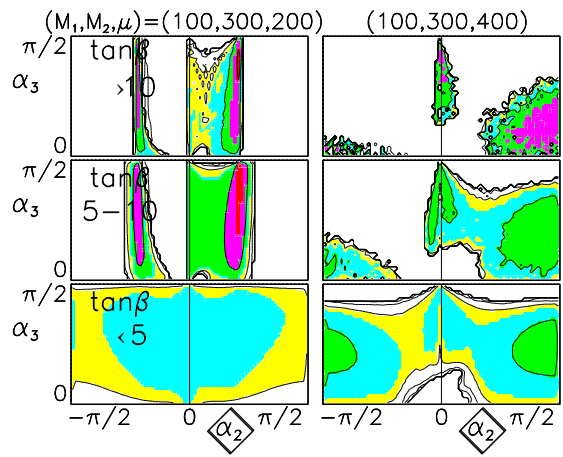}
\caption{\label{Fig:alphas-300-23}
Allowed regions in the $\alpha_2$--$\alpha_3$ plane, for
$(M_1,M_2)=(100,300)~\GeV$, $\mu=200$~GeV and 400~GeV
and three slices in $\tan\beta$ as indicated.
Contours and colour coding as in Fig.~\ref{Fig:alphas-cont2-300-500}.}
\end{center}
\vspace*{-2mm}
\end{figure}

\subsection{$(M_1,M_2)=(100, 300)~\GeV$}

We display in Figs.~\ref{Fig:alphas-300} and \ref{Fig:alphas-300-23} the
allowed regions in the $\alpha_1$--$\alpha_2$ and $\alpha_2$--$\alpha_3$
planes, for three slices in $\tan\beta$ and two values of $\mu$, ``small''
(200~GeV) and ``moderate'' (400~GeV).  At $\tan\beta<5$
sizable regions in $\vecalpha$ space are populated with allowed solutions, but
these regions shrink significantly for $\tan\beta>5$.

From Figs.~\ref{Fig:alphas-300} and \ref{Fig:alphas-300-23}, we see that for
large $\tan\beta$ and increasing $\mu$, the majority of solutions have values
of $\alpha_2$ that move away from 0 towards $\pi/2$ (where CP is conserved,
see Fig.~\ref{Fig:alpha23-scheme}).
In this case of increasing $\mu$, $M_3$ will also increase, and the last two
terms of (\ref{Eq:lambda1}) must compensate each other.  In distinction from
the case of $\mu=0$, $|s_1|$ and $|s_2|$ can then not both be small, one of
them will approach unity, as seen from Fig.~\ref{Fig:alphas-300}.  According
to (\ref{Eq:positivity}), $\lambda_3$ can not be too large and negative. From
Eq.~(\ref{Eq:lambda3}), this means that either $s_1$ (i.e., $\alpha_1$) or
$s_2$ (i.e., $\alpha_2$) must be large and positive, as seen in
Figs.~\ref{Fig:alphas-300} and \ref{Fig:alphas-300-23}.

There will for large values of $\tan\beta$ be a range of $\mu$-values for
which the allowed solutions accumulate around $\alpha_2\simeq\pm\pi/4$.  In
these regions, the CP-violation in the $H_1b\bar b$-coupling will be
considerable.

The distribution of $M_3$-values is shown in Table~\ref{tab:mh3-dist-100-300}.
Here, we note that for $M_1<\mu<M_2$, the values of $M_3$ are rather low
(close to $M_2$), whereas for $M_2<\mu$ they tend to be considerably higher.

\begin{table}[htb]
\begin{center}
\renewcommand{\tabcolsep}{.75em}
$(M_1,M_2,\mu)=(100,300,200\ [400])$~GeV
\begin{tabular}{|c|c|c|c|}
\hline 
$\tan\beta$&$\xi<1.1$&$1.1<\xi<1.5$&$1.5<\xi$\\
\hline
$>10$ &58.6 [\ 0.0]\%&40.0 [70.4]\%& 1.4 [29.6]\%\\
$5-10$&41.8 [\ 0.0]\%&55.3 [44.2]\%& 3.0 [55.8]\%\\
$<5$  &29.7 [\ 0.0]\%&56.1 [13.1]\%&14.2 [86.9]\%\\
\hline
\end{tabular} 
\caption{Distribution of $M_3$ values, $\xi=M_3/M_2$.}
\label{tab:mh3-dist-100-300}
\end{center}
\vspace*{-2mm}
\end{table}

\subsection{$(M_1,M_2)=(100, 500)~\GeV$}

\begin{figure}[t!]
\begin{center}
\vspace*{-2mm}
\includegraphics[width=9.cm]{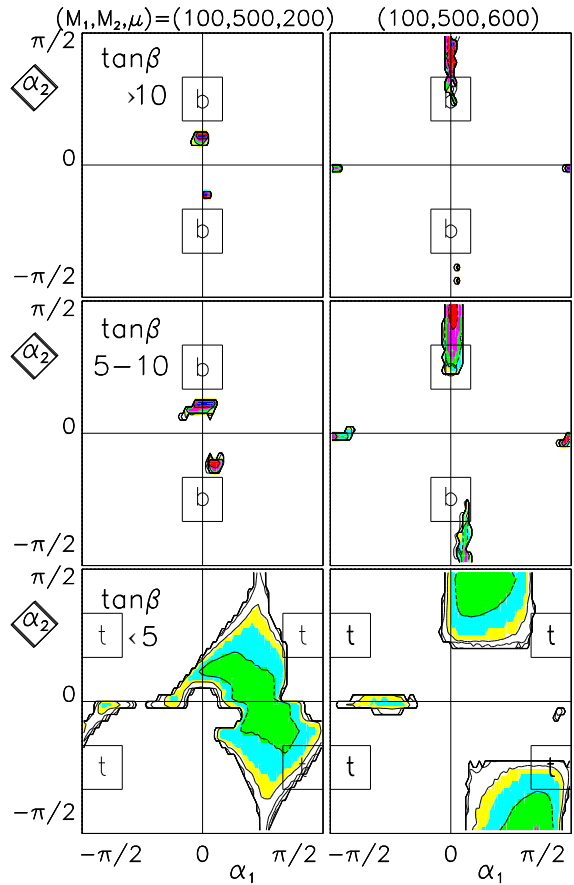}
\caption{\label{Fig:alphas-500}
Allowed regions in the $\alpha_1$--$\alpha_2$ plane, for
$(M_1,M_2)=(100,500)~\GeV$, $\mu=200$~GeV and 400~GeV
and three slices in $\tan\beta$ as indicated.  Regions of major CP violation
are labeled ``$b$'' and ``$t$''.}
\end{center}
\vspace*{-2mm}
\end{figure}

For this case of ``large'' $M_2$, we display in Fig.~\ref{Fig:alphas-500} the
allowed regions in the $\alpha_1$--$\alpha_2$ plane, for three slices in
$\tan\beta$ and two values of $\mu$, 200~GeV and 600~GeV.

In this region of large $M_2$, we note that at high values of $\tan\beta$, the
allowed regions in $\alpha_1$ get constrained to values around
$\alpha_1\simeq0$ or $\alpha_1\simeq\pm\pi/2$ with $|\alpha_2|$ increasing
with $\mu$ from 0 to $\pi/2$.  (We recall that the $\alpha_2\to\pi/2$ limit
represents the case when the lightest Higgs particle, $H_1$, is odd, and there
is no CP violation.)  It follows from Eq.~(\ref{Eq:M_3}) that when
$|\alpha_1|$ and $|\alpha_2|$ are both small, $M_3$ is close to
$M_2$. Furthermore, it follows from (\ref{Eq:lambda1}) that in this limit, we
have
\begin{equation} \label{Eq:largetanbeta}
\lambda_1\simeq\frac{1}{c_\beta^2 v^2}
[\half M_1^2+\half s_3^2 M_2^2 +\half c_3^2 M_3^2 - s_\beta^2\mu^2].
\end{equation}
This should not get too large, in order not to spoil unitarity.  Since $M_2$
and $M_3$ are comparable in this case, the distribution in $\alpha_3$ becomes
wide. This is analogous to the situation shown in
Fig.~\ref{Fig:alphas-300-23}, left panel.

\begin{table}[htb]
\begin{center}
\vspace*{2mm}
\renewcommand{\tabcolsep}{.75em}
$(M_1,M_2,\mu)=(100,500,200/ [600])$~GeV
\begin{tabular}{|c|c|c|c|}
\hline 
$\tan\beta$&$\xi<1.1$&$1.1<\xi<1.5$&$1.5<\xi$\\
\hline
$>10$&93.7 [0.0]\%&\ 6.3 [94.8]\%&0.0 [5.2]\%\\
$5-10$&84.9 [0.0]\%&15.1 [93.7]\%&0.0 [6.3]\%\\
$<5$&74.1 [0.0]\%&25.9 [91.5]\%&0.0 [8.5]\%\\
\hline
\end{tabular} 
\caption{Distribution of $M_3$ values, $\xi=M_3/M_2$.}
\label{tab:mh3-dist-100-500}
\end{center}
\vspace*{-2mm}
\end{table}

As discussed above, when $\tan\beta\gg1$, the allowed range of $M_3$
values get squeezed to a narrow band just above $M_2$. This is illustrated
in Table~\ref{tab:mh3-dist-100-500}.

When $\mu$ increases still, the situation becomes reminiscent of that shown in
Fig.~\ref{Fig:alphas-300}, right part. The majority of allowed solutions move
towards $\alpha_1\simeq0$ and $\alpha_2\simeq\pm\pi/2$, with small islands of
additional solutions at $\alpha_1\simeq\pm\pi/2$ and $\alpha_2\simeq0$.

\section{CP-conserving limits}\label{sec:CP-conservation}
\setcounter{equation}{0}

In addition to the general criteria for limits of no CP violation given in
Eq.~(\ref{Eq:CP-cons}) and Fig.~\ref{Fig:alpha23-scheme}, there are
important limits in which there is no CP violation in the Yukawa couplings
involving the {\it lightest} Higgs boson: $b\bar bH_1$ and $t\bar tH_1$.

\subsection{CP-conserving $b\bar bH_1$ coupling}
The regions of CP-invariant $b\bar bH_1$ coupling require {\it either}
$R_{11}\simeq0$ (implying $\alpha_1\simeq\pm\pi/2$ and/or
$\alpha_2\simeq\pm\pi/2$) {\it or} $R_{13}\simeq0$ (implying
$\alpha_2\simeq0$). These limits are shown in Fig.~\ref{Fig:alpha12-scheme}.
We see from Figs.~\ref{Fig:alphas-cont2-300-500}, \ref{Fig:alphas-300} and
\ref{Fig:alphas-500} that both these categories of CP-conserving regions exist
for $\mu=0$ as well as for $\mu>0$.

\subsection{CP-conserving $t\bar tH_1$ coupling}
\begin{figure}[hbt]
\vspace*{-5mm}
\begin{center}
\includegraphics[width=7cm]{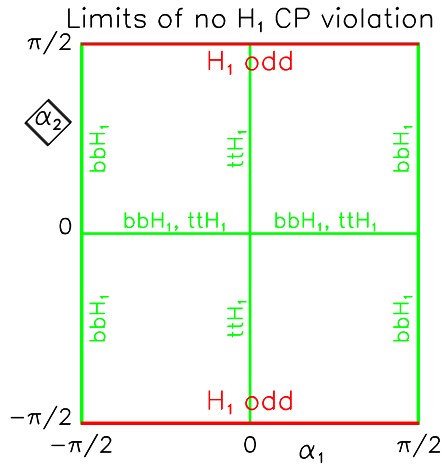}
\vspace*{-5mm}
\caption{\label{Fig:alpha12-scheme} Limits of no CP-violation in the 
$b\bar bH_1$ and $t\bar tH_1$ Yukawa couplings, shown in the 
$\alpha_1$--$\alpha_2$ plane. Also indicated, are the limits of $H_1$ odd,
where there is no CP violation involving {\it any} of the three Higgs bosons.}
\end{center}
\vspace*{-2mm}
\end{figure}

The regions of CP-invariant $t\bar tH_1$ coupling require {\it either}
$R_{12}\simeq0$ (implying $\alpha_1\simeq0$ and/or $\alpha_2\simeq\pm\pi/2$)
{\it or} $R_{13}\simeq0$ (implying $\alpha_2\simeq0$).  These limits are shown
in Fig.~\ref{Fig:alpha12-scheme}.  We see from the lower panels in
Figs.~\ref{Fig:alphas-cont2-300-500}, \ref{Fig:alphas-300} and
\ref{Fig:alphas-500} that such regions exist for $\mu=0$ as well as for
$\mu>0$.

\section{Two light Higgs bosons}
\label{sect:two-light}
\setcounter{equation}{0}
In this section, we report on some results obtained with $M_1=100$~GeV and
$M_2=150$~GeV (or 200~GeV). There are two questions: (1) Which parts of the
$\tan\beta$--$M_{H^\pm}$ plane are populated by allowed solutions, and (2) to
what extent do such models provide CP violation?  Concerning the latter
question, we recall that two light Higgs bosons will to some extent act
coherently, with Yukawa strength proportional (by orthogonality) to that
of the heaviest one, $H_3$.
\subsection{$\mu=0$}
This case is similar to the case discussed in Sec.~\ref{sect:mu=0} in the
sense that for $\mu=0$ (or small), there is an upper limit to $\tan\beta$.
That limit is lifted as $\mu$ is increased.  At low $\tan\beta$ and $\mu=0$,
for example, the majority of solutions fall in the domain of small
$|\alpha_1|$, small $|\alpha_2|$, and large $\alpha_3$, as is required for
significant CP violation in the $t$-quark sector (see the discussion at the
end of Sec.~\ref{sec:CP-violation}).  At larger values of $\tan\beta$ (and
$\mu=0$), the solutions still populate the small $|\alpha_1|$, small
$|\alpha_2|$ region, which is unfavorable for CP violation in the $b$-quark
sector.
\subsection{$M_2<\mu$}
Let us first consider the case of $\tan\beta={\cal O}(1)$. As compared with
the case $\mu=0$, when $\mu$ increases, $|\alpha_1|$, $|\alpha_2|$ 
and  $|\alpha_3|$ all
tend to move towards larger values, $|\alpha_1|\to\pi/2$, $|\alpha_2|\to\pi/2$,
$\alpha_3\to\pi/2$. This limit does not satisfy the conditions
for major CP violation in the $t$-quark sector.
At larger values of $\tan\beta$, the solutions move
towards intermediate and negative values of $\alpha_1$, 
intermediate values of $\alpha_2$, and intermediate values of $\alpha_3$,
which is a favorable parameter region for CP violation in the $b$-quark sector.
\section{Concluding remarks}\label{sec:summary}
\setcounter{equation}{0}

We have made a survey of parameter regions of large CP violation in the Two
Higgs Doublet Model.  Because of the many independent model parameters, it is
difficult to extract a simple picture.  For the admittedly limited set of
parameters studied, it was found that considerable CP violation can easily
occur in the $t\bar tH_1$ coupling at low values of $\tan\beta$.  In order to
have significant CP violation in the $b\bar bH_1$ coupling, on the other hand,
it appears necessary to have a large mass splitting among 
the two heavier Higgs bosons, $H_2$ and $H_3$.
\bigskip

\medskip
\leftline{\bf Acknowledgment} 
\par\noindent 
It is a pleasure to thank O. Brein, M. Krawczyk and A. Pilaftsis for very 
useful discussions.  This research has been supported in part by the Mission
Department of Egypt and by the Research Council of Norway.

\goodbreak

\end{document}